\documentclass[]{aa}
\usepackage{graphicx}
\begin{document}
\title{CCD photometric search for peculiar stars in open clusters. VIII. 
King~21, NGC~3293, NGC~5999, NGC~6802, NGC~6830, Ruprecht~44, Ruprecht~115, and Ruprecht~120 
\thanks{Based on 
observations at CASLEO, CTIO (Proposal
2003A-0057), and OSN. The Observatorio de Sierra Nevada is operated by the
Consejo Superior de Investigaciones Cient\'ificas through
the Instituto de Astrof\'isica de Andaluc\'ia (Granada, Spain)}}
\author{M.~Netopil\inst{1}, E.~Paunzen\inst{1}, H.M.~Maitzen\inst{1},
O.I.~Pintado\inst{2,}\thanks{Visiting Astronomer at Complejo Astron\'omico El 
Leoncito operated under an agreement between Consejo Nacional de 
Investigaciones Cient\'ificas y T\'ecnicas de la Rep\'ublica Argentina 
and the National Universities of La Plata, Cordoba, and San Juan.},
A.~Claret\inst{3}, L.F.~Miranda\inst{3}, I.Kh.~Iliev\inst{4}, V.~Casanova\inst{3}}

\mail{Martin.Netopil@univie.ac.at}

\institute{Institut f\"ur Astronomie der Universit\"at Wien,
           T\"urkenschanzstr. 17, A-1180 Wien, Austria
\and	   Departamento de F\'isica, Facultad de Ciencias Exactas 
           y Tecnolog\'ia, Universidad Nacional de Tucum\'an, Argentina - Consejo Nacional 
		   de Investigaciones Cient\'ificas y T\'ecnicas de la Rep\'ublica Argentina
\and	   Instituto de Astrof\'isica de Andaluc\'ia
		   CSIC, Apartado 3004, 18080 Granada, Spain
\and	   Institute of Astronomy, National Astronomical Observatory, 
           P.O. Box 136, BG-4700 Smolyan, Bulgaria}

\date{Received 2006 / Accepted 2006}
\authorrunning{M. Netopil et al.}
\titlerunning{Photometric search for peculiar stars in open clusters. VIII.}

\abstract{We continue our survey for magnetic chemically peculiar (CP2)
stars in galactic open clusters to shed more light on their origin and evolution.}
{To study the group of CP2 stars, it is essential to find these objects
in different galactic environments and at a wide range of evolutionary stages. 
The knowledge of open clusters ages and metallicities can help to find a
correlation of these parameters with the (non-)presence of peculiarities 
which has to be taken into account in stellar evolution models.}
{The intermediate band $\Delta a$ photometric system 
samples the depth of the 5200\AA\, flux depression by comparing the flux 
at the center with the adjacent regions with bandwidths of 110\AA\, to 230\AA. 
It is capable to detect magnetic CP2 and CP4 stars with high
efficiency, but also the groups of (metal-weak) $\lambda$ Bootis, 
as well as classical Be/shell stars can be successfully investigated.
In addition, it allows to determine the age, reddening and
distance modulus with an appropriate accuracy by fitting isochrones.}
{From the 1677 observed members of the eight open clusters, 
twenty five CP2 and one Ae stars were identified. Further nineteen deviating stars are 
designated as questionable due to several reasons. The estimated 
age, reddening and distance for the programme clusters were compared with
published vales of the literature and discussed in this context.}
{The current paper shows that CP2 stars are present continuously 
in very young (7\,Myr) to intermediate age (500\,Myr) open clusters at distances
larger than 2\,kpc from the Sun.}

\maketitle

\section{Introduction}

In continuation of our CCD $\Delta a$ photometric survey to detect 
chemically peculiar (CP) stars of the upper main sequence, eight more  
open clusters were chosen, including very young (log\,$t$\,=\,6.8) to old (log\,$t$\,=\,8.7) aggregates lying at distances up to 4.8\,kpc from the sun. 

Sampling their characteristic flux depression at 5200\,\AA$ $ (cf. Kupka et al. \cite{kup03}, \cite{kup04}), the $\Delta a$ photometric system 
is capable of detecting CP and related objects in an easy and economic way. Paunzen et al. (\cite{pau05a}) 
re-examined recently the efficiency of this system, resulting in a detection probability of up to 95\,\% for all relevant magnetic CP stars. Furthermore, the groups of $\lambda$Bootis and classical Be/shell stars systematically exhibit negative $\Delta a$ values.

The formation and evolution of CP stars is still a matter of debate; unambiguous detection or 
even non-detection of those groups in different galactic environments will therefore help to shed more light on that topic. The latest study of the incidence of peculiar objects was carried out by North (1993) on the basis of 57 open clusters, which is still a small number considering the quantity of $\sim$ 1700 known clusters in the Milky Way (Dias et al. \cite{dia02}). The $\Delta a$ system is therefore the best solution for extending the examined sample. It also serves as a kind of pre-selection for further studies. Recently, Bagnulo et al. (\cite{bag06}) for example used $\Delta a$ results to obtain spectropolarimetry at the ESO VLT.  

However, if one aims to cover clusters at different galactic environments from the solar neighbourhood, a lack of deeper investigations, especially of membership analysis, becomes evident, necessary for 
segregating possible peculiar stars among nonmembers. We are therefore planning to obtain additional UBV photometry, wherever applicable, during future observations.

Beside detecting CP stars, we are also able to determine cluster parameters such as age, 
reddening, and distance, using isochrones for the $\Delta a$ system (Claret et al. \cite{cla03}, 
Claret \cite{cla04}), which were compared with already published parameters. Since Paunzen \& Netopil (\cite{pau06a}) show that the accuracy of parameters for a 
lot of open clusters is still far from satisfactory, this additional capability of $\Delta a$ is very important. More data sets will contribute to minimising the errors for individual clusters, which then helps for examining the galactic structure more accurately, among other possibilities.     

In the currently presented sample of eight galactic clusters (King~21, NGC~3293, NGC~5999, NGC~6802, NGC~6830,  
Ruprecht~44, Ruprecht~115, and Ruprecht~120) we detected 
25 CP2, 1 Ae, and 19 objects showing peculiar behaviour due to doubtful membership, variability, or binarity. 

\section{Observations, reduction, and used methods}

Observations of the eight open clusters were performed 
at three different sites and telescopes: 
\begin{itemize}
\item 0.9\,m telescope (CTIO), direct imaging, SITe 2084\,$\times$\,2046 pixel CCD,  
13$\arcmin$ field-of-view
\item 2.15\,m RC telescope (CASLEO), direct imaging, EEV CCD36-40 1340\,$\times$\,1300 pixel 
CCD, 9$\arcmin$ field-of-view
\item 1.5\,m RC telescope (OSN), direct imaging, RoperScientific VersArray 2048\,$\times$\,2048 pixel 
CCD, 8$\arcmin$ field-of-view.
\end{itemize}
The observing log with the number of frames in each filter
is listed in Table \ref{log}. The observations were performed with
two different filter sets, both with the following characteristics:
$g_1$ ($\lambda_c$\,=\,5007\,\AA, FWHM\,=\,126\,\AA, $T_P$\,=\,78\%), 
$g_2$ (5199, 95, 68) and $g_3$ (5466, 108, 70), whereby $T_P$ indicates the transmission efficiency.

The basic CCD reductions and a point-spread-function fitting 
were carried out within standard IRAF V2.12.2 routines on
personal computers running under LINUX.
The way of calculating the normality line, deriving the errors,
the calibration of our $(g_1-y)$, and $y$ measurements, 
is the same as in previous works (see Netopil et al. \cite{net05} and Paunzen et al. \cite{pau06b} for example) 
and will not be repeated here in detail. Our measured
$y$ magnitudes, as well as the $(g_1-y)$ colour indices were directly converted into standard $UBV$ magnitudes
on the basis of already published values. The transformation coefficients can be found in Table \ref{coeffs}.
 
\begin{table*}[t]
\begin{center}
\caption{Observing log for the programme clusters with the number of frames in each filter. 
The exposure time per frame is given in seconds.}
\label{log}
\begin{tabular}{lllcll}
\hline\hline
Cluster & Site & Date & \#${g_{\rm 1}}$/${g_{\rm 2}}$/${y}$ & Exp.time & Observer\\
\hline
King~21 & OSN & August 28, 2005 & 10/10/10 & 180/360 & V. Casanova\\
NGC~3293 & CTIO & April 21, 2003 & 10/10/10 & 5/10 & H.M. Maitzen\\
NGC~5999 & CTIO & April 21$-$23, 2003 & 10/10/10 & 300 & H.M. Maitzen\\
NGC~6802 & OSN & August 22, 2005 & 10/10/10 & 180/360 & V. Casanova\\
NGC~6830 & OSN & August 24, 2005 & 10/10/10 & 180/360 & V. Casanova\\
Ruprecht~44 & CTIO & April 22, 2003 & 5/5/5 & 100/200 & H.M. Maitzen\\
Ruprecht~115 & CASLEO & June 26$-$27, 2003 & 10/10/10 & 300 & O.I. Pintado\\
Ruprecht~120 & CTIO & April 22$-$23, 2003 & 9/9/9 & 300 & H.M. Maitzen\\
\hline
\end{tabular}
\end{center}
\end{table*}

\begin{figure*}
\begin{center}
\includegraphics[width=155mm]{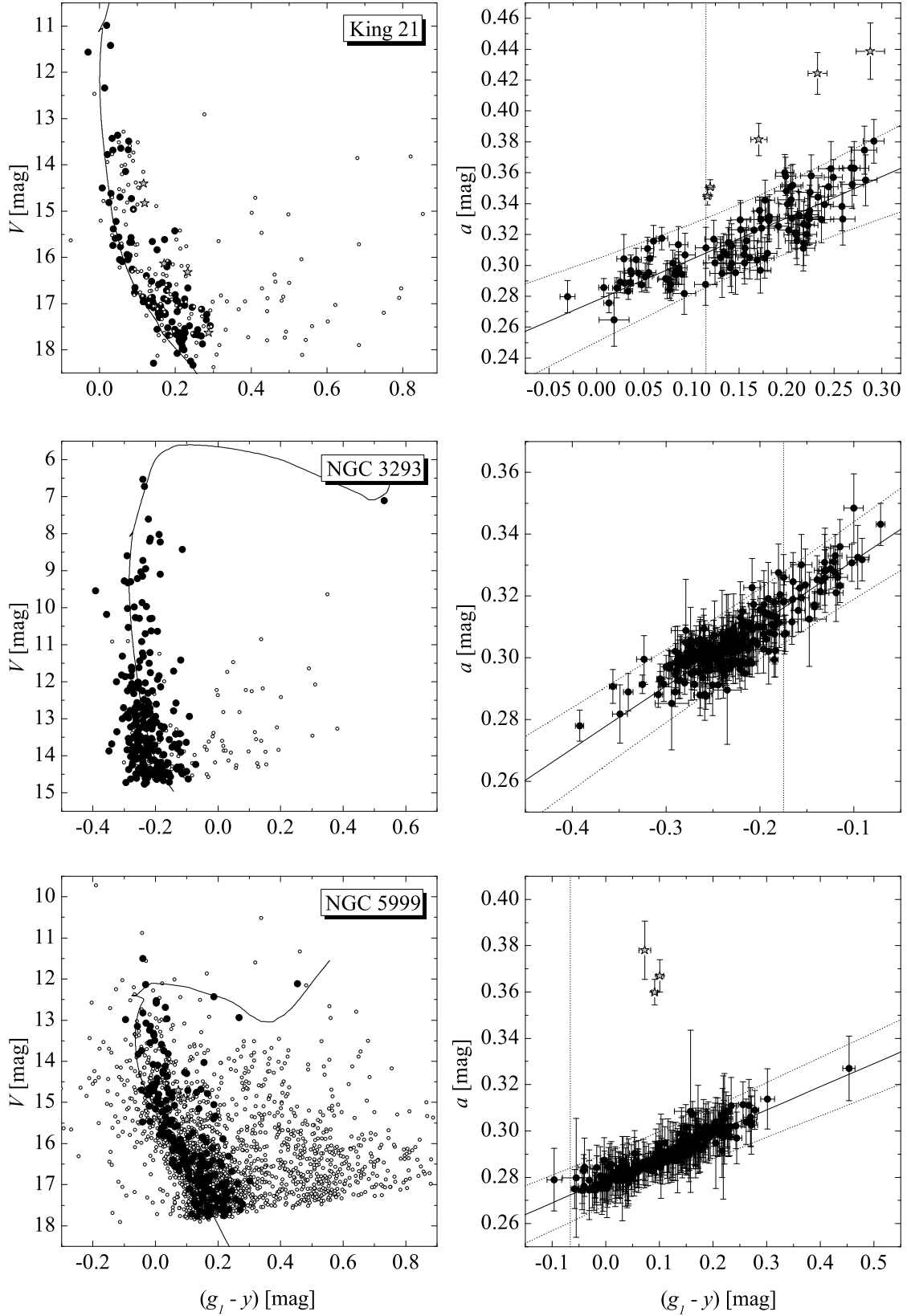}
\caption[]{Observed $a$ versus $(g_{1}-y)$ and $V$ versus $(g_{1}-y)$
diagrams for our programme clusters. Left panels: The solid lines represent the corresponding 
isochrones taken from Claret et al. (\cite{cla03}) and Claret (\cite{cla04}), which are based on the
$\Delta a$ photometric system. The derived ages, reddenings, and
distance moduli are given in Table \ref{all_res}. Right panels: The solid line is the
normality line, whereas the dotted lines are the confidence intervals
corresponding to 99.9\,\%. The error bars for each individual object
are mean errors. The detected peculiar objects are marked by
stars. Only members (filled circles) have been used to derive the
normality lines, their coefficients can be found in Table \ref{coeffs}. 
The dotted vertical lines represent the colour index for spectral type A0. }
\label{fig1}
\end{center}
\end{figure*}

\begin{figure*}
\begin{center}
\includegraphics[width=155mm]{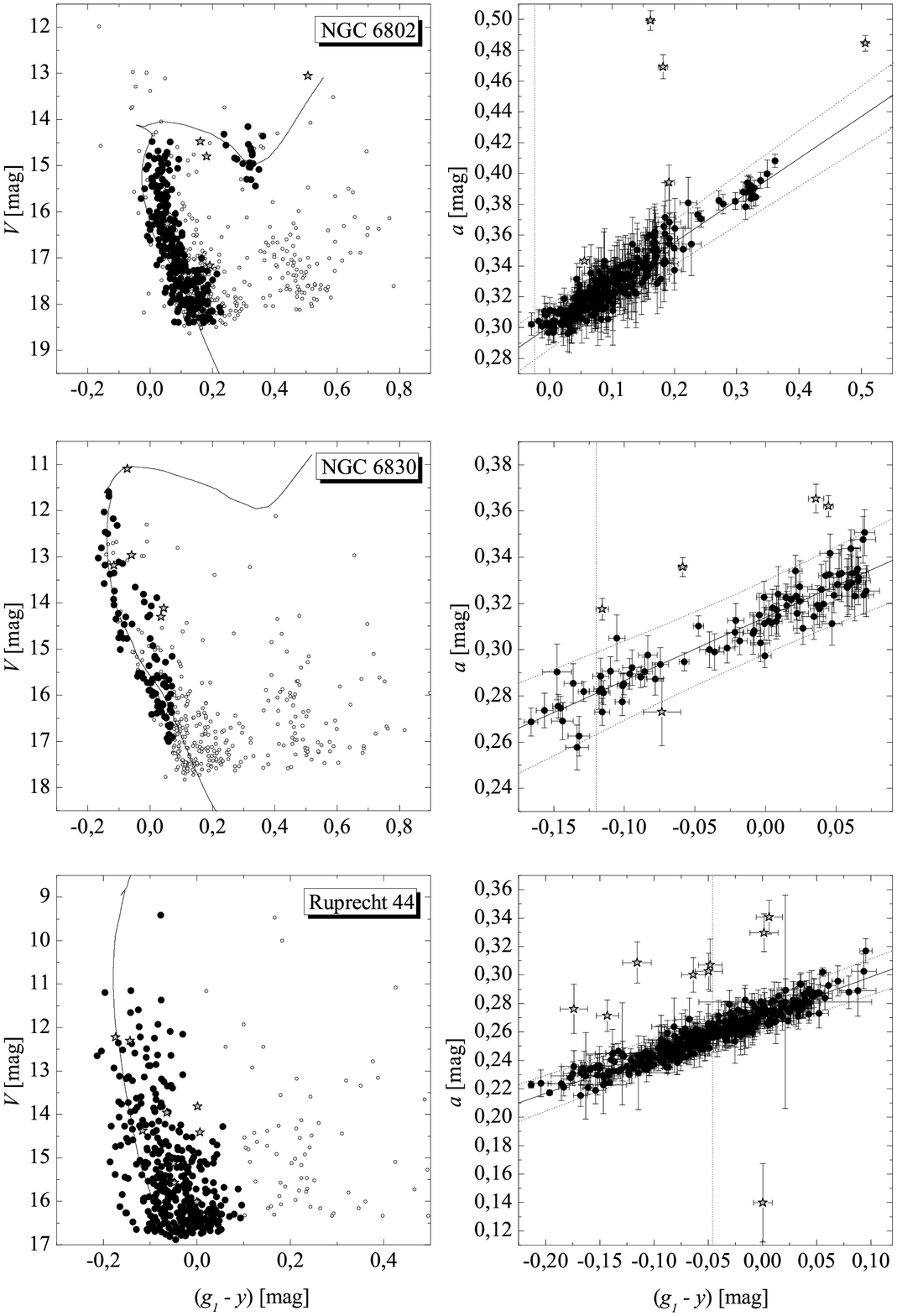}
\caption[]{Observed $a$ versus $(g_{1}-y)$ and $V$ versus $(g_{1}-y)$
diagrams for our programme clusters. The symbols
are the same as in Fig. \ref{fig1}.}
\label{fig2}
\end{center}
\end{figure*}

\begin{figure*}
\begin{center}
\includegraphics[width=155mm]{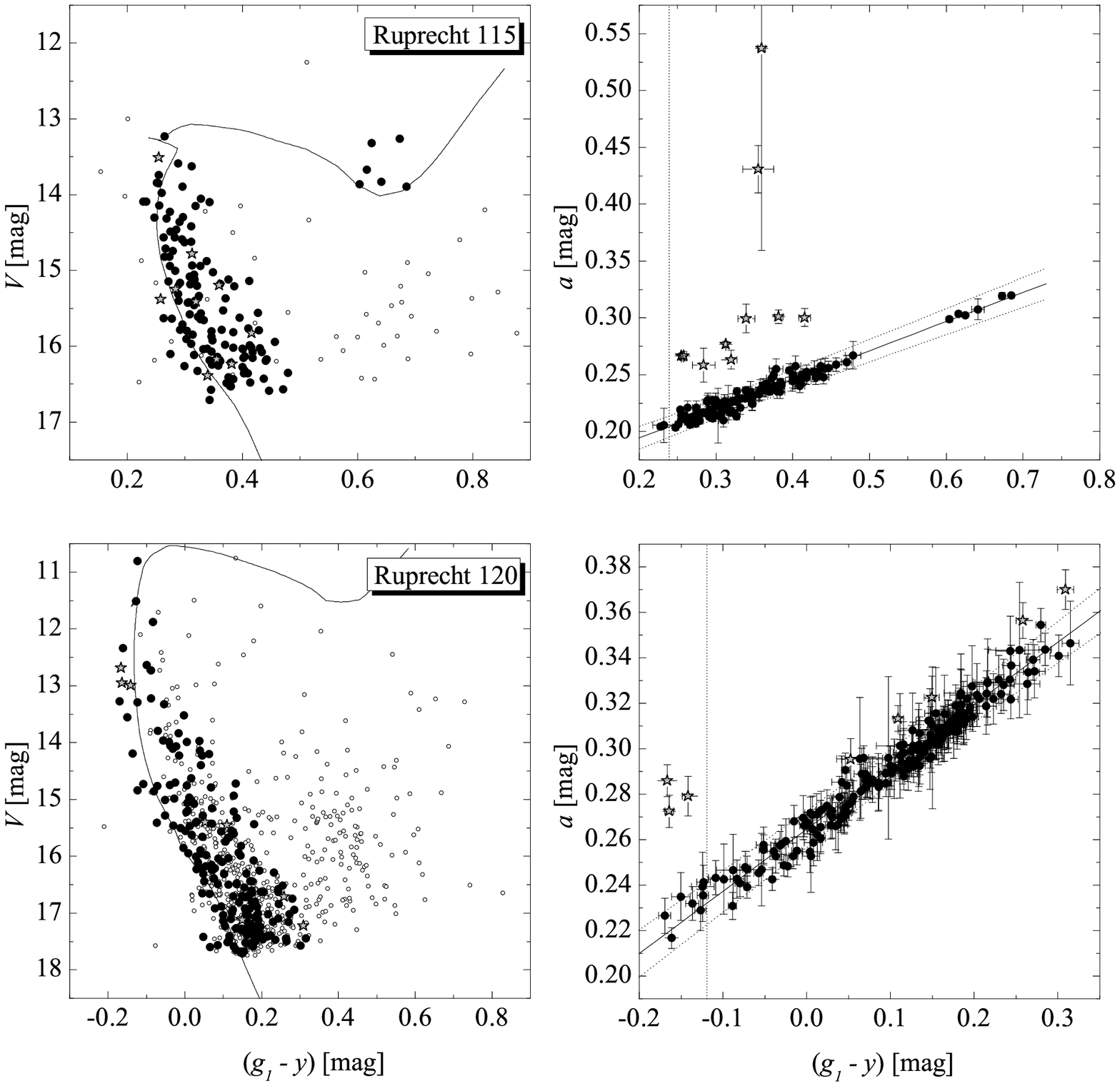}
\caption[]{Observed $a$ versus $(g_{1}-y)$ and $V$ versus $(g_{1}-y)$
diagrams for our programme clusters. The symbols
are the same as in Fig. \ref{fig1}.}
\label{fig3}
\end{center}
\end{figure*}

The isochrones shown in Figs. \ref{fig1} to \ref{fig3} were taken from
Claret et al. (\cite{cla03}) and Claret (\cite{cla04}), which are based on the
$\Delta a$ photometric system. The derived ages, reddenings, and
distance moduli, together with their errors are listed in Table \ref{all_res}.
The fitting procedure takes advantage of the available $UBV$ measurements for
all programme clusters, which means that the results were compared to
those of the colour-magnitude diagrams for the $UBV$ photometric system.
However, our determination is based on the $\Delta a$ measurements
alone, which is another important application of this photometric system.

\begin{table*}
\begin{minipage}[t]{\textwidth}

\caption{Summary of results: the age, distance modulus, reddening,
and thus the distance from the Sun derived by fitting isochrones
to the $\Delta a$ photometry.   
}

\label{all_res}
\centering
\renewcommand{\footnoterule}{}
\scriptsize{
\begin{tabular}{llll}
\hline\hline
Name & King~21 & NGC~3293 & NGC~5999  \\
     & \object{C2347+624} & \object{C1033$-$579} & \object{C1548$-$563}  \\
\hline
$l/b$ & 115.95/+0.68  &  285.86/+0.07  & 326.01/$-$1.93     \\
$E(B-V)$ ($\pm$0.05) & 0.85 & 0.29  & 0.48      \\
$m_V - M_V$ ($\pm$0.2) & 15.3 & 13.1  & 13.2      \\
$d$\,[kpc] & 3.41(56)\footnote{The errors of the last digits are always given in parenthesis.}   & 2.76(45)  & 2.20(36)    \\
$R_{GC}$\,[kpc] & 9.98(38)  & 7.72(2)  & 6.30(26)   \\
$|z|$\,[pc] & 41(7) & 4(1) & 74(12) \\
log\,$t$ ($\pm$0.1) & 7.2 & 7.0  & 8.5  \\
Tr-type\footnote{The Trumpler classification was taken 
from Lyng{\aa} (\cite{lyn87}).} & I 2 p & I 3 r & I 2 m  \\
n(member) & 115 & 241 & 227    \\
n(none)   & 144 & 96  & 1960 \\   
n(frames) & 30  & 30  & 30    \\
CP No. (Webda):            & 55($-$):+59/+113/+832      &   &  1570(119): +92/+223/+1509     \\
$\Delta$a/$(B-V)_0$/$M_V$  & 85($-$):+85/+242/+1020     &   & 1642(148): +72/+421/+1721     \\
$[$mmag$]$                 & 137($-$):+41/+6/$-$465     &   & 1693($-$): +78/+405/+2217      \\
                           & 151($-$):+84/+358/+2335    &   &                               \\
                           & *152(11)\footnote{Questionable peculiar stars are marked by asterisks.}:+36/0/$-$895      &   &                                \\
\hline\hline
Name & NGC~6802 & NGC~6830 & Ruprecht~44  \\
     & \object{C1928+201} & \object{C1948+229} & \object{C0757$-$284}  \\
\hline
$l/b$ & 55.33/+0.92 & 60.14/$-$1.80  & 245.75/+0.48  \\
$E(B-V)$ ($\pm$0.05) & 0.89 & 0.48  & 0.58    \\
$m_V - M_V$ ($\pm$0.2) & 14.4 & 12.9  & 15.2    \\
$d$\,[kpc] & 2.13(35) & 1.92(31) & 4.79(79)  \\
$R_{GC}$\,[kpc] & 7.01(13) & 7.24(10) & 10.88(57)   \\
$|z|$\,[pc] & 34(6) & 60(10) & 40(7)  \\
log\,$t$ ($\pm$0.1) & 8.7 & 8.3 & 6.8  \\
Tr-type & I 1 m & II 2 p & IV 2 m \\
n(member) & 279 & 92 & 379 \\
n(none)   & 401 & 311 & 66  \\   
n(frames) & 30 & 30  & 15    \\
CP No. (Webda):           & *208($-$): +41/+635/+2769 & 117(38): +38/+255/+50       & 16($-$): +68/+149/$-$784       \\
$\Delta$a/$(B-V)_0$/$M_V$  &  *314(42): +46/+1360/$-$1470 & *164(9): $-$21/+170/$-$1900 & *17(235): +39/$-$370/$-$2680 \\
$[$mmag$]$                 & *345(100): +155/+230/+90 & *175(180): +42/+365/+1395   & *143(162): +52/$-$380/$-$2900  \\
                           & *395(138): +119/+460/+480  & 257(166): +35/$-$65/+290    & 381($-$): +69/$-$197/$-$825  \\
                           & 427($-$): +27/+233/+1870 & *324(159): +36/+485/+1245   & 383($-$): $-$132/+133/+814     \\
                           &        &                     & 394($-$): +45/$-$9/$-$70       \\
                           &        &                     & 436($-$): +49/$-$6/+369        \\
                           &        &                     & 443($-$): +46/$-$49/$-$1235    \\
                           &        &                     & 444($-$): +58/+136/$-$1380     \\
                           
\hline\hline
Name & Ruprecht~115 & Ruprecht~120 & \\
     & \object{C1609$-$522} & \object{C1631$-$482} & \\
\hline
$l/b$ &  330.96/$-$0.85  & 336.39/$-$0.49 &\\
$E(B-V)$ & 0.74  & 0.65  &\\
$m_V - M_V$ ($\pm$0.2) & 13.8  & 13.5  &\\
$d$\,[kpc] & 2.00(33) & 1.98(33) &\\
$R_{GC}$\,[kpc] & 6.33(26) & 6.24(28)&  \\
$|z|$\,[pc] & 30(5) & 17(3)& \\
log\,$t$ ($\pm$0.1) & 8.6  & 8.0& \\
Tr-type &  III 1 p & II 3 p&\\
n(member) & 142 & 202&\\
n(none)   & 50  & 549&\\   
n(frames) & 30  & 27 & \\
CP No. (Webda):                & *54(30): +302/+487/+1484  & 485(34): +53/$-$49/$-$411 &\\
$\Delta$a/$(B-V)_0$/$M_V$      & *58(93): +60/+514/+2422    & 487(35): +52/$-$51/$-$536 &\\
$[$mmag$]$                    & 65(63): +38/+297/+1618    & *490($-$): +17/+387/+1906&\\
                               & *73(3): +197/+430/+2095   & 600(38): +66/$-$49/$-$769&\\
                                & 81(38): +43/+208/+1464    & *604($-$): +22/+970/+3728&\\
                           & 90($-$): +53/+206/+980    & *616($-$): +18/+607/+3623&\\
                           &  92(7): +58/+64/$-$282     & *623($-$): +22/+854/+3209&\\
                           & 98(11): +57/+79/1564      & *663(116): +19/+480/+1980&\\
                            & 132(178): +69/+431/2610   & &\\
                           &  *139(57): +51/+639/2029    & &\\
  \hline

\end{tabular} }
\end{minipage}
\end{table*}

The tables with all the data for the cluster stars, as well as
nonmembers, are available
in electronic form at the CDS via anonymous ftp to cdsarc.u-strasbg.fr (130.79.128.5),
http://cdsweb.u-strasbg.fr/Abstract.html, at WEBDA (http://www.univie.ac.at/webda/)
or upon request from the first author. These tables include the cross
identification of objects from the literature, the $X$ and $Y$
coordinates within our frames, the observed $(g_{1}-y)$ and
$a$ values with their corresponding errors, $V$ magnitudes,
the $(B-V)$ colours from the literature, as well as the 
$\Delta a$-values derived from the normality lines based on $(g_{1}-y)$
(disregarding nonmembers).

The diagnostic diagrams for
all eight open clusters are shown in Figs. \ref{fig1} to \ref{fig3}. Furthermore,
the normality lines and the confidence intervals corresponding to 99.9\,\%
are plotted. The detected peculiar objects are marked by
stars. Only members have been used to derive the
normality lines. Membership was attributed according to their
location in the colour-magnitude diagrams, the distance from the
cluster centres, and additional information
from the literature (proper motions and radial velocities) taken from
WEBDA.

\section{Results} \label{results}

\begin{table*}
\begin{minipage}[t]{\textwidth}
\caption{Comparison of our results (indicated by bold face) to already published cluster 
parameters.}
\label{comparison}
\centering
\renewcommand{\footnoterule}{}
\begin{tabular}{lllll}
\hline\hline
Cluster & log\,$t$ & $E(B-V)$ & $d$\,[kpc] & Ref   \\
\hline
\textbf{King~21}  & \textbf{7.20}(1)\footnote{The errors of the last digits are always given in parenthesis.} & \textbf{0.85}(5) & \textbf{3.41}(56) & \\
                  &      & 0.89  & 1.91 & Mohan \& Pandey (1984)              \\
                  & 7.16 & 0.89  & 2.10 & Dias et al. (2002)               \\

\textbf{N~3293}   & \textbf{7.00}(1) & \textbf{0.29}(5) & \textbf{2.76}(45) &\\
		          & 6.70 & 0.33 & 2.50 & Turner et al. (1980)                  \\
		          & 7.00 & 0.31 & 2.60 & Feinstein \& Marraco (1980)          \\
				  & 7.00 & 0.25\footnote{Interstellar reddening was transformed according $E(b-y)$\,=\,0.74$E(B-V)$ (Crawford \cite{cra78}).} & 2.75 & Balona (1994)                      \\
		          & 6.90 & 0.29 & 2.75 & Baume et al. (2003)                  \\
		          & 6.94 & 0.25 & 2.47 & Kharchenko et al. (2005)              \\
                  & 7.01 & 0.26    & 2.33 & Dias et al. (2002)                \\

\textbf{N~5999}   & \textbf{8.50}(1) & \textbf{0.48}(5) & \textbf{2.20}(36) & \\
                  & 8.00 & 0.44 &      & Santos \& Bica (1993)                \\
                  & 8.60 & 0.45 & 2.20 & Piatti et al. (1999)                  \\
                  & 8.60 & 0.45 & 2.05 & Dias et al. (2002)                    \\
                  
\textbf{N~6802}   & \textbf{8.70}(1) & \textbf{0.89}(5) & \textbf{2.13}(35) & \\
				          &      & 0.83 & 0.91 & Hoag \& Applequist (1965)    \\
				          &      & 0.80 & 0.82:\footnote{Indicates an uncertain result in the respective reference.}& Becker \& Fenkart (1971)     \\  
				          & 9.18:& 0.80 & 1.84 & Kaluzny \& Shara (1988)      \\ 
                  & 9.00 & 0.94 & 1.43 & Sirbaugh et al. (1995)               \\
                  & 8.87 & 0.85 & 1.12 & Dias et al. (2002)                 \\
                  
 \textbf{N~6830}   & \textbf{8.30}(1) & \textbf{0.48}(5) & \textbf{1.92}(31) &  \\        	                    	          &       &      & 1.68 & Barkhatova (1957) \\		         
   		 &       & 0.51 & 1.38 & Hoag \& Applequist (1965) \\ 
   		  	     &       & 0.58 & 1.47 & Becker \& Fenkart (1971)\\
        &  8.00 & 0.56 & 1.70 & Moffat (1972)\\
         &  7.52 & 0.50 & 1.64 & Kharchenko et al. (2005)\\
         &  7.57 & 0.50 & 1.64 & Dias et al. (2002)\\
 \textbf{Ru~44}    & \textbf{6.80}(1) & \textbf{0.58}(5) & \textbf{4.79}(79) &    \\
  &        & 0.64 & 4.60 & Turner (1981)  \\
   &        & 0.68 & 6.80 & FitzGerald \& Moffat (1976)\\
   &  6.00  & 0.70 & 6.60 & Moffat \& Fitzgerald (1974)  \\
&  7.33  & 0.62 & 4.66 & Kharchenko et al. (2005)  \\
&  6.94  & 0.62 & 4.73 & Dias et al. (2002)  \\
 \textbf{Ru~115}   & \textbf{8.60}(1) & \textbf{0.74}(5) & \textbf{2.00}(33) &    \\
   & 8.70 & 0.60 & 2.50 & Piatti et al. (1999)\\
   & 8.78 & 0.65 & 2.16 & Dias et al. (2002)\\
 \textbf{Ru~120}   & \textbf{8.00}(1) & \textbf{0.65}(5) & \textbf{1.98}(33) &  \\
    & 8.00 & 0.65  & 2.30 & Piatti et al. (1999)\\
    & 8.18 & 0.70  & 2.00 & Dias et al. (2002)\\

\hline
\end{tabular}
\end{minipage}
\end{table*}

In the following we present the results for the individual open clusters and compare them with
the literature. The cluster parameters and their errors are listed 
in Tables \ref{all_res} and \ref{comparison}, and the  
significant deviating cases are discussed in more detail. The ratio of total-to-selective extinction $R_{V}$ was set to 3.1, and for the distance of the Sun from
the galactic centre a value of 8.0\,kpc was used. Averaged $(B-V)$ and $V$ values 
were taken from the literature, corrected by the obtained distance modulus and interstellar reddening. If not available, we used the transformed $\Delta a$ photometry (see Table \ref{coeffs}).  
Star numbers are based on ascending (X,Y) coordinates
on the CCD frames. In Table \ref{results}, one can also find 
the star numbers according to WEBDA, if available. 

\subsection{King~21} \label{KI21}

The only available photometric study about King~21 was performed by Mohan \& Pandey (\cite{moh84}). 
They measured 26 stars in the cluster area only down to about 14$\fm$5, biasing the 
determination of the distance modulus. This explains the large difference 
in distance compared to our result, as is seen in Table \ref{comparison}. Furthermore, they estimated the cluster age according to the method by Mermilliod (\cite{mer81}), settling it between the very young groups of NGC~2362 and NGC~457. This would correspond to an age around log\,$t$=7.0. Five possible chemically peculiar stars (\#55, 85, 137, 151, 152) with $\Delta a$ indices between +36 and +85\,mmag were found within this aggregate, all lying in the typical spectral range of CP2 stars, according to their colour indices. However, star \#152 has to be treated with caution, since it is part of a close pair. Together with object \#155, it represents the star \#11 according to Mohan \& Pandey (\cite{moh84}). Some other stars are lying somewhat outside the determined confidence interval, but are considered insignificant due to their errors. The authors of the foregoing study report a nonuniform reddening across the cluster, which could be responsible for the location of the five detected 
peculiar stars in the colour-magnitude diagram: all objects deviate slightly from the cluster main sequence. Thus, a definite conclusion about their membership cannot be made yet due to the scarcity of information. Moreover, we are unable to confirm the variability of object \#94 reported by Mohan \& Pandey (\cite{moh84}) (\#19 therein), because of our shorter observing period. 

\subsection{NGC~3293} \label{NGC3293}

Due to its brightness, several studies were performed on the young cluster NGC~3293, including an extensive spectroscopic survey by Evans et al. (\cite{eva05}) using the telescopes VLT (Flames) and ESO 2.2m (FEROS). Its location in the area of the Carina Nebula (NGC~3372) enhances its attractivity.
Although the detection limit of our $\Delta a$ photometry is very low for that cluster, we did not find any peculiar object. However, several stars showing H$\alpha$ emission have already been identified in the former investigations by Baume et al. (\cite{bau03}) and McSwain \& Gies (\cite{mcs05}). Furthermore, the survey by Evans et al. (\cite{eva05}) resulted in the detection of five Be/Ae objects, 
as well as one definite Si star and an object showing peculiar He I lines. The Si star is located outside of our field-of-view and is a probable nonmember due to its large distance from the cluster centre, whereas Evans et al. (\cite{eva05}) note for the other object (\#22 according to our photometry, $\Delta a$ = +3\,mmag) that the peculiar appearance is probably due to a hotter, though less luminous, companion. 
Astonishing is the non visibility of the large number of likely Be/Ae objects in $\Delta a$. Inspecting the indicated emission stars of the three studies mentioned above, one can notice that, among the nine targets in common, no detection coincides in all investigations. Just two stars (WEBDA\,\# 12 \& 117) were designated as emission objects in at least two studies. These stars exhibit $\Delta a$ indices of $-$8 and $-1$\,mmag respectively.
McSwain \& Gies (\cite{mcs05}), for example, note that their results are likely contaminated by the background H$\alpha$ emission, as well as emission from pre-main sequence stars. Furthermore, the phases of emission can be replaced by shell and normal phases of the object. Such behaviour was investigated e.g. by Pavlovski \& Maitzen (\cite{pav89}) using $\Delta a$ photometry. The determined cluster parameters are in perfect agreement with the literature values (Table \ref{comparison}). For the sake of a better presentation of the diagnostic $\Delta a$ diagram (Fig. \ref{fig1}), only main-sequence stars were plotted. Nevertheless, the single red giant shown in the colour-magnitude diagram of NGC~3293 in Fig. \ref{fig1} is found close to the normality line.

\subsection{NGC~5999} \label{NGC5999}

Beside the search for variable stars within the OGLE survey (Pietrzynski et al. \cite{pie98}), the only in-depth 
investigation of the open cluster NGC~5999 was carried out by 
Piatti et al. (\cite{pia99}). Their results are in perfect agreement with the parameters obtained in this work, in contrast 
to Santos \& Bica (\cite{san93}) who found a much younger age (log\,$t$\,=\,8.0) based on integrated spectra and template 
matching. Additionally, they also obtained a slightly higher age of $\sim$ 230\,Myr using Balmer lines 
and a metallicity of [Fe/H]=0.18 from Ca\,II triplet lines. However, we assumed solar metallicity 
for the isochrone-fitting procedure, since the more recent work by Santos \& Piatti (\cite{san04}) list [Fe/H]=0.0(2). 
Although apparent red giants seem to be present 
in the innermost region of NGC~5999, the obtained parameters are based only on main sequence stars, because 
neither their spectral type nor their membership is known. 
Also the two brightest ``main sequence'' stars of our sample, \#1326 (WEBDA \#36) and \#1465 (WEBDA \#53), were disregarded for determining the cluster parameters, since we have obtained much redder colours (up to $\sim$\,0$\fm$2) compared to Piatti et al. (\cite{pia99}) 
and their $(B-V)$ and $(V-I)$ colours, whereas the $V$ magnitudes agree with previous work. 
A reinvestigation of these stars is necessary in order to determine their cluster affiliation. 
If they turn out to be members, a much younger age has to be set for the cluster. Because of the small cluster diameter compared to the available field-of-view and the high star density in that field, a lot of nonmembers have to be rejected for further analysis (see Table \ref{all_res}, Fig. \ref{fig1}).
The detection of three extremely deviating 
objects can be reported (\#1570, 1642, and 1693). They exhibit $\Delta a$ indices between +72 and +92\,mmag and are, due 
to their colours, early/mid F type stars, just on the cool end of the spectral range where 
CP stars are expected. Since no membership analysis is available 
in the literature, the only hint can be delivered by the present colour-magnitude diagram, 
which does not contradict membership, although they do deviate slightly from the main sequence.

\subsection{NGC~6802} \label{NGC6802}

Within the oldest cluster in the current sample, we were able to detect five possibly peculiar stars, 
but no classical ones except for one (\#427). Although that star exhibits a moderate $\Delta a$ 
index of 27\,mmag, it lies just slightly outside the confidence interval due to the large 
scatter among the cluster members. One other suspicious star (\#208) is, on the one hand, too 
cool to be a CP2 object; on the other hand, it clearly deviates from the main sequence, 
suggesting a questionable membership status. All other stars are placed along the giant branch, 
but their membership is uncertain. Evolved objects of luminosity classes III to I can also show the 
same $\Delta a$ behaviour as CP2. In addition, two of them (\#345, 395) exhibit exceptional $\Delta a$ indices 
of more than 100\,mmag. The clear visible giant clump in the colour-magnitude diagram 
allows the cluster parameters to be determined more accurately, but the obtained distance of 2.13\,kpc 
is in part considerably higher than in the literature. By inspecting the photometry 
by Hoag et al. (\cite{hoa61}), one can notice numerous brighter stars in an extended area around NGC~6802 that also form a cluster main sequence. Including them in an analysis will result in a much lower distance. 
However, it is up to later observations to clarify whether those brighter stars mimic a cluster sequence or build a foreground cluster.

\subsection{NGC~6830} \label{NGC6830}
According to earlier investigations, ages between log\,$t$=7.5 and 8.0 were found for the cluster NGC~6830, 
whereas we have obtained a much older age of log\,$t$=8.3, which represents the best fit to the 
current data. Grubissich (\cite{gru60}) notes, that the brightest cluster stars are probably not physical members. This objects are not measured by us, because of saturation or lying outside of our field-of-view.
If they are included in an analysis, a much 
younger age will be derived, of course. The above-mentioned stars (\#2, 5, and 72 according to WEBDA) have spectral types of B9 IIIe, B7 V, and B9.5 III (Hoag \& Applequist \cite{hoa65}), respectively. The authors determined a true distance module of 10$\fm$7 based on photometric $UBV$, H$\gamma$, and 
spectrographic observations of six bright cluster stars in total, which is 0$\fm$7 lower than 
the one obtained by us on the basis of the whole main sequence. This seems to confirm the assumption 
by Grubissich (\cite{gru60}). Due to a large scatter of the cooler stars within the $\Delta a$ diagram, 
probably caused by nonmembers and/or the photometric errors of the fainter stars, they were 
truncated at about $(B-V)$=1.0, since CP stars are no longer expected at cooler temperatures. 
Five deviating objects were found in the remaining domain, whereas the only negative star (\#164) 
is questionable due to its larger error, but the spectral type of A9 III by Turner (\cite{tur76}) 
did not contradict its membership status. Two of the four objects showing a positive $\Delta a$ index 
(\#175 \& 324) have an adopted spectral type of mid F and are already at the cool end of CP stars, 
exhibiting the largest deviation from the main sequence, whereas the remaining two objects 
(\#117 \& 257) are candidates for classical, chemically-peculiar stars. One star (\#226, WEBDA \#24) 
of our sample is also included in the catalogue of H$\alpha$ emission-line stars by 
Kohoutek \& Wehmeyer (\cite{koh99}), showing a $\Delta a$ index of $-$16\,mmag, just lying within 
the confidence interval. Restricting ourselves to the hotter portion ($(g_{1}-y)$\,$<$\,$-$0.05), the calculation of the normality line and its confidence interval would yield an outlying position below it.

\subsection{Ruprecht~44} \label{Ru44}
Ruprecht~44 is another very young cluster in the current survey. However, it is most difficult 
to fit isochrones to data of such young aggregates, because of the nonexistence of a giant clump or 
a turn-off point to determine an accurate distance module. Due to ongoing star formation, differential 
reddening can be an additional difficulty. We therefore used only the innermost region to 
determine the cluster parameters, which well agree with the literature values. 
Only Moffat \& FitzGerald (\cite{mof74}) and FitzGerald \& Moffat (\cite{fit76}) obtained much larger distances. These discrepancies already have been discussed by Turner (1981).		 
Nine possible peculiar objects were found within this cluster, which are all well separated in the 
$\Delta a$ diagram. There is only one star (\#383) that exhibits a huge negative $\Delta a$ 
index ($-$132\,mmag), whereas its photometric index implies an Ae nature. Unfortunately, 
a spectroscopic confirmation will be challenging due to its close position to the bright 
star \#383 ($V$= 9$\fm$48). Furthermore, the objects \#17 (A7:) and 143 (B8:) are nonmembers 
according to Moffat \& FitzGerald (\cite{mof74}).
All other stars are magnetically peculiar 
candidates with $\Delta a$ indices between 45 and 69\,mmag. We have to note that, although 
the cluster is extended ($\sim$ 12$\arcmin$ in diameter), filling nearly the whole available 
field-of-view, most of the detected peculiar stars are lying at the assumed cluster border. Another 
interesting behaviour is the concentration of the probable CP2 objects, and four of them are located 
in an area of about 2$\times$2$\arcmin$. On the basis of these findings, a reinvestigation of 
the whole $\Delta a$ survey will be necessary to answer the question of whether local fluctuations 
are possibly affecting the formation of the CP phenomenon.   

\subsection{Ruprecht~115} \label{Ru115}
The cluster parameters obtained for Ruprecht~115 partly agree with results found in 
the literature. Piatti et al. (\cite{pia99}) have studied the cluster via $BVI$ photometry and integrated 
spectra. They have determined an age that is only slightly older than our value. The differences in the 
results for interstellar reddening E($B-V$) can be explained 
by their combined values for the determinations via photometry and integrated spectra, whereas they 
have obtained a comparable value via photometry (0.65), but a lower one (0.50) using the spectra 
template method. However, when inspecting their reddening 
result E($V-I$)=0.9 and the ratio E($V-I$)/E($B-V$)=1.25 by Dean et al. (\cite{dea78}), our result of E($B-V$)=0.74 
seems to be in good agreement, assuming a normal extinction law. The stars in the apparent red giant clump 
are probable members, since they all lie directly on the normality line, suggesting a similar 
reddening. In total, ten possible peculiar objects were found, two of them (\#54 and 73) showing extreme 
$\Delta a$ indices of +302 and +197\,mmag, respectively. These are rather unusual values by implying nonmembership 
or more probably variability, since a comparison of our calibrated $V$ and ($B-V$) magnitudes with the 
previous investigation exhibits differences up to $\sim$ 0$\fm$3, but further investigations are necessary. 
One of them (\#54) also has a large error, mainly caused by the results for the $g_{2}$ filter, which 
covers the 5200\AA$ $ flux depression. These objects, together with two other stars (\#58 and 139) that are too cool to be classified as CP members, were therefore marked as questionable in Table \ref{all_res}.
The six other positive detections exhibit $\Delta a$ indices in the 
``normal'' range between +38 and +69\,mmag, as well as colour indices hotter than spectral type F5. 
However, a membership analysis 
is not available for the detected objects, but none (also the extreme cases) deviates in the colour-magnitude diagram.  

\subsection{Ruprecht~120} \label{Ru120}
We have found eight possibly peculiar objects within the cluster Ruprecht~120. The 
photometric indices for three of them (\# 485, 487, 600) suggest late B-type nature. 
The range of their positive $\Delta a$ indices (52 to 66\,mmag) is indicative of their 
bona fide CP2 nature (Paunzen et al. \cite{pau05a}). The remaining five CP candidates have 
moderate $\Delta a$ indices up to +22\,mmag, but are too cool for classical CP objects, 
except \#490, an early/mid F-type star, lying just on the border of the confidence interval,  
therefore designated as questionable. The obtained cluster parameters agree with 
the work by Piatti et al. (\cite{pia99}), but they have determined a slightly greater distance. 
A lot of field stars, designated as nonmembers due to their distance to the cluster 
centre (a diameter of 7$\arcmin$ was assumed), show a similar behaviour to the 
cluster main sequence. Such an appearance of galactic field stars was already 
discussed for example by Piatti et al. (1999).  

\begin{table}[t]
\begin{minipage}[t]{\columnwidth}
\caption{The regression coefficients for the
transformations and normality lines.}

\label{coeffs}
\centering
\renewcommand{\footnoterule}{}
\begin{tabular}{l|l|l}
\hline\hline
Cluster & $V$, $(B-V)$, $a_{0}$\footnote{The absolute values and errors
vary due to the inhomogeneous ``standard'' observations (photographic, photoelectric
and CCD) found in the literature as well as the dependence on the 
magnitude range in common, i.e. a broader range guarantees a small error.  
The errors of the last digits are given in parenthesis.}  & N  \\
\hline 
King~21 & $-$0.42(16)\,+\,1.00(1)$\cdot y$ & 19 \\
        & 0.61(1)+2.09(24)$\cdot(g_{1}-y)$ & 19    \\
        & 0.277(3)\,+\,0.267(16)$\cdot(g_{1}-y)$ & 110 \\
NGC~3293  & $-$5.31(8)\,+\,1.00(1)$\cdot y$ & 72 \\
        & 0.75(3)+2.63(11)$\cdot(g_{1}-y)$ & 67 \\
        & 0.352(1)\,+\,0.203(5)$\cdot(g_{1}-y)$ & 241 \\
NGC~5999 & $-$5.35(6)\,+\,1.00(1)$\cdot y$ & 278  \\
        & 0.64(1)+2.42(5)$\cdot(g_{1}-y)$ & 273  \\
        & 0.279(1)\,+\,0.100(3)$\cdot(g_{1}-y)$ & 222  \\
NGC~6802 & $-$0.15(19)\,+\,1.06(1)$\cdot y$ & 98 \\
        & 0.96(1)+2.96(9)$\cdot(g_{1}-y)$ & 92  \\
        & 0.301(1)\,+\,0.273(6)$\cdot(g_{1}-y)$ & 274  \\
NGC~6830 & 0.93(15)\,+\,0.98(1)$\cdot y$  & 35 \\
        & 0.81(1)+2.76(7)$\cdot(g_{1}-y)$ & 35  \\
        & 0.314(1)\,+\,0.274(11)$\cdot(g_{1}-y)$ & 87  \\
Ruprecht~44 & $-$5.32(10)\,+\,1.00(1)$\cdot y$ & 75 \\
        & 0.71(1)+2.84(12)$\cdot(g_{1}-y)$ & 75  \\
        & 0.271(1)\,+\,0.273(1)$\cdot(g_{1}-y)$ & 370  \\
Ruprecht~115 & $-$4.30(7)\,+\,1.00(1)$\cdot y$  & 72  \\
        & 0.08(3)+2.76(7)$\cdot(g_{1}-y)$ & 72  \\
        & 0.143(2)\,+\,0.256(5)$\cdot(g_{1}-y)$ & 132  \\
Ruprecht~120 & $-$6.38(9)\,+\,1.05(1)$\cdot y$  & 93  \\
        & 0.92(1)+2.27(6)$\cdot(g_{1}-y)$ & 93  \\
        & 0.265(1)\,+\,0.270(4)$\cdot(g_{1}-y)$ & 194  \\
\hline

\end{tabular}
\end{minipage}

\end{table}

\section{Conclusions}

\begin{figure}
\begin{center}
\includegraphics[width=90mm]{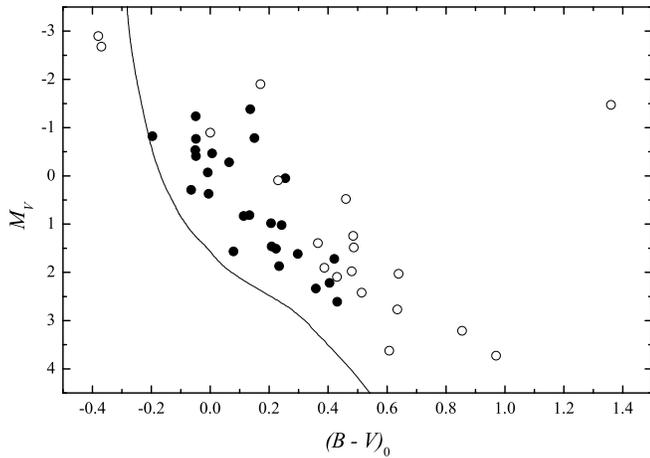}
\caption[]{The detected probable (filled circles) and questionable (open circles) peculiar stars 
within the $(B-V)_{0}$/$M_{V}$ plane as given in Table \ref{all_res}. The solid line represents 
the ZAMS relation from Schmidt-Kaler (\cite{sch82}).}
\label{fig4}
\end{center}
\end{figure}

Within the presented sample of eight galactic open clusters, 5254 objects (1677 cluster members) on 222 frames using 
three different sites were investigated, resulting in the detection of twenty five CP2 stars and one 
object of a probable Ae nature. Another nineteen deviating stars are 
designated as questionable for to several reasons (e.g. uncertain membership status) as indicated in the respective 
cluster section. Due to the poor available membership information, we also studied 
their behaviour in the $(B-V)_{0}$/$M_{V}$ diagram (Fig. \ref{fig4}) using the values given in 
Table \ref{all_res}. The star closest to the ZAMS is object \#381 in Ruprecht~44, 
the youngest aggregate of the current sample. No other distinctive features can be found if taking the differential reddening of some clusters into account, but the most 
outstanding objects are \#117 (NGC~6830), as well as \#16 and 44 (Ruprecht~44), which exhibit the largest 
deviations from the ZAMS.  

As an important application of the $\Delta a$ photometric system, 
isochrones were fitted to the colour-magnitude diagrams ($V$ versus
$(g_1-y)$) of the programme clusters. A comparison of our results
yields excellent agreement with the corresponding parameters from
the literature. 

The programme clusters are all located at distances $\geq$\,2\,kpc from the sun, widely distributed in 
galactocentric distances (6.2 to 10.9\,kpc) and ages (6 to 500\,Myr), thus helping to cover different 
environments and evolutionary stages for a deeper investigation of the incidence of peculiar stars. 
Of special interest is the influence of metallicity, taking its gradients within the 
galactic disc into account (Chen et al. \cite{che03}). Together with NGC~3105 and Berkeley~11 of earlier $\Delta a$ investigations (Paunzen et al. \cite{pau05b}; Paunzen et al. \cite{pau06b}), the cluster Ruprecht~44 in 
the current sample is farthest from the galactic centre ($\sim$10.7\,kpc), applying 
a value of $R_{0}=8.0$. We have to note that the $R_{GC}$ data for NGC~3105 in Paunzen et al. (\cite{pau05b}) 
was wrongly given as 11.4 instead of 11.0 using $R_{0}=8.5$. Examining their CP stars content in respect to the investigated total members, lead to 1.6, 2.1 and 5.6\% for Ruprecht~44, NGC~3105, and Berkeley~11,  respectively, neglecting questionable candidates. However, only Ruprecht~44 and NGC~3105 represent clusters of comparable 
age (log\,$t$=6.8 and 7.3) and galactic coordinates, whereas Berkeley~11 deviates by age (log\,$t$\,=\,8.0) and position. On account of the small number, we want to enhance our efforts with clusters situated in outer regions, but also in the opposite direction, since the solar neighbourhood has been investigated up to now a lot via photoelectric and CCD $\Delta a$ photometry as well as spectroscopy. 

\begin{acknowledgements}
This research was performed within the projects  
{\sl P17580} and {\sl P17920} of the Austrian Fonds zur F{\"o}rderung der 
wissen\-schaft\-lichen
Forschung (FwF) and also benefited from financial contributions from the City of
Vienna (Hochschuljubil{\"a}umsstiftung project: H-1123/2002). 
I.Kh.~Iliev acknowledges 
the support by the Bulgarian National Science Fund under grant F-1403/2004 and the Pinehill Foundation. 
Use was made of the SIMBAD database, operated at the CDS, Strasbourg, France and
the WEBDA database, operated at the University
of Vienna. This research made use of NASA's Astrophysics Data System.
\end{acknowledgements}


\begin{thebibliography}{}
\bibitem[2006]{bag06} Bagnulo, S., Landstreet, J. D., Mason, E., et al. 2006, \aap, 450, 777
\bibitem[1994]{bal94} Balona, L. A. 1994, \mnras, 267, 1060
\bibitem[1957]{bar57} Barkhatova, K. A. 1957, \sovast, 1, 822
\bibitem[2003]{bau03} Baume, G., V\'azquez, R. A., Carraro, G., \& Feinstein, A. 2003, \aap, 402, 549
\bibitem[1971]{bec71} Becker, W., \& Fenkart, R. 1971, \aaps, 4, 241
\bibitem[2003]{che03} Chen, L., Hou, J. L., \& Wang, J. J. 2003, \aj, 125, 1397
\bibitem[2004]{cla04} Claret, A. 2004, \aap, 424, 919
\bibitem[2003]{cla03} Claret, A., Paunzen, E., \& Maitzen, H. M. 2003, \aap, 412, 91
\bibitem[1978]{cra78} Crawford, D. L. 1978, \aj, 83, 48
\bibitem[1978]{dea78} Dean, J. F., Warren, P. R., \& Cousins, A. W. J. 1978, \mnras, 183, 569
\bibitem[2002]{dia02} Dias, W. S., Alessi, B. S., Moitinho, A., et al. 2002, \aap, 389, 871
\bibitem[2005]{eva05} Evans, C. J., Smartt, S. J., Lee, J.-K., et al. 2005, \aap, 437, 467
\bibitem[1980]{fei80} Feinstein, A., \& Marraco, H. G. 1980, \pasp, 92, 266 
\bibitem[1976]{fit76} FitzGerald, M. P., \& Moffat, A. F. J. 1976, \aap, 50, 149
\bibitem[1994]{god94} Goderya, S. N., \& Schmidt, E. G. 1994, \apj, 426, 159
\bibitem[1960]{gru60} Grubissich, C. 1960, \zap, 50, 14
\bibitem[1965]{hoa65} Hoag, A. A., \& Applequist, N. L. 1965, \apjs, 12, 215
\bibitem[1961]{hoa61} Hoag, A. A., Johnson, H. L., Iriarte, B., et al. 1961, Publ. U. S. nav. Obs., 2nd Ser., vol. XVII, Part VII
\bibitem[1988]{kal88} Kaluzny, J., \& Shara, M. M. 1988, \aj, 95, 785
\bibitem[2005]{kha05} Kharchenko, N. V., Piskunov, A. E., R\"oser, S., Schilbach, E., \& Scholz, R.-D. 2005, \aap, 438, 1163
\bibitem[1999]{koh99} Kohoutek, L., \& Wehmeyer, R. 1999, \aaps, 135, 255
\bibitem[2003]{kup03} Kupka, F., Paunzen, E., \& Maitzen, H. M. 2003, \mnras, 341, 849
\bibitem[2004]{kup04} Kupka, F., Paunzen, E., Iliev, I. Kh., et al. 2004, \mnras, 352, 863
\bibitem[1987]{lyn87} Lyng\aa\,\,G. 1987, Catalogue of Open Cluster Data, 5$^{\rm th}$ edition, CDS, Strasbourg
\bibitem[2005]{mcs05} McSwain, M. V., \& Gies, D. R. 2005, \apjs, 161, 118
\bibitem[1981]{mer81} Mermilliod, J. C. 1981, \aap, 97, 235
\bibitem[1972]{mof72} Moffat, A. F. J. 1972, \aaps, 7, 355
\bibitem[1974]{mof74} Moffat, A. F. J., \& FitzGerald, M. P. 1974, \aap, 34, 291
\bibitem[1984]{moh84} Mohan, V., \& Pandey, A. K. 1984, BASI, 12, 217
\bibitem[2005]{net05} Netopil, M., Paunzen, E., Maitzen, H. M., et al. 2005, Astron. Nachr., 326, 734
\bibitem[1993]{nor93} North, P. 1993, PASPC, 44, 577
\bibitem[2006]{pau06a} Paunzen, E., \& Netopil, M. 2006, \mnras, 371, 1641
\bibitem[2005a]{pau05a} Paunzen, E., St\"utz, Ch., \& Maitzen, H. M. 2005a, \aap, 441, 631
\bibitem[2005b]{pau05b} Paunzen, E., Netopil, M., Iliev, I. Kh., et al. 2005b, \aap, 443, 157
\bibitem[2006]{pau06b} Paunzen, E., Netopil, M., Iliev, I. Kh., et al. 2006, \aap, 454, 171
\bibitem[1989]{pav89} Pavlovski \& Maitzen 1989, \aaps, 77, 351
\bibitem[1992]{pet92} Petry, C. E., \& DeGioia-Eastwood, K. 1992, \baas, 24, 1235
\bibitem[1999]{pia99} Piatti, A. E., Claria, J. J., \& Bica E. 1999, \mnras, 303, 65
\bibitem[1998]{pie98} Pietrzynski, G., Kubiak, M., Udalski, A., \& Szymanski, M. 1998, \actaa, 48, 498
\bibitem[1993]{san93} Santos, J. F. C. Jr., \& Bica, E. 1993, \mnras, 260, 915
\bibitem[2004]{san04} Santos, J. F. C. Jr., \& Piatti, A. E. 2004, \aap, 428, 79
\bibitem[1982]{sch82} Schmidt-Kaler, Th. 1982, in Landolt-B{\"o}rnstein New Series,
Group VI, Vol. 2b, p. 453
\bibitem[1995]{sir95} Sirbaugh, R. D., Lewis, K. A., \& Friel, E. D. 1995, \baas, 27, 1439
\bibitem[1976]{tur76} Turner, D. G. 1976, \aj, 81, 1125
\bibitem[1981]{tur81} Turner, D. G. 1981, \aj, 86, 222
\bibitem[1980]{tur80} Turner, D. G., Grieve, G. R., Herbst, W., \& Harris, W. E. 1980, \aj, 85, 1193
\end{thebibliography}
\end{document}